\documentclass[12pt]{article}

\usepackage{graphics}
\usepackage{float}
\usepackage{epsfig}
\usepackage{epstopdf}
\usepackage{setspace}
\usepackage{amssymb}

\title{Holographic Ricci dark energy and generalized second law}

\author{Titus K Mathew and Praseetha P \\
Department of Physics, Cochin University of Science and Technology, \\ Kochi-682022, India. \\ 
E-mail:titus@cusat.ac.in and praseethapankunni@gmail.com.}

\date{}

\begin{document}

\maketitle

\begin{abstract}
We explore the validity of the generalized second law (GSL) of thermodynamics in flat FRW universe with apparent horizon and event 
horizon as the boundary. We found that in a universe with holographic Ricci dark energy and dark matter, interacting with each other, 
 the GSL is satisfied at the 
apparent horizon and partially satisfied at the event horizon under thermal equilibrium conditions. We also analyses the GSL under 
non-equilibrium conditions and shows that the fulfillment of GSL at the apparent horizon implies that the temperature of the dark 
energy is greater than that of the horizon. Thus there occurs a flow of dark energy towards the apparent horizon. As a result the 
entropy of the dark energy decreases and that of horizon increases. This is verified by finding the evolution of the dark energy 
entropy and horizon entropy in a dark energy dominated universe under non-equilibrium conditions.

\end{abstract}
\hspace{0.33in} Keywords:Dark energy, generalized second law, FRW universe.

\section{Introduction}
\label{intro}

Observations on type Ia supernova have shown that the present universe is dominated by dark energy with negative pressure which
provides a mechanism for the accelerating expansion of the universe \cite{PM1,AG1,Garno1}. Cosmic Microwave Background anisotropies also 
supporting the cosmic acceleration \cite{Spergel1}. Various theoretical models of dark energy are proposed and are mainly concentrating 
on predicting the strength of this cosmic acceleration. Most of these models are based on scalar or multi scalar fields, e.g. quintessence 
model \cite{Ratra1,Wetterich1,Zlatev1}, K-essence model \cite{Chiba1}, Chaplygin gas model \cite{Kamenshchik1, Alam1}, phantom model 
\cite{Caldwel1}, quintom model \cite{Feng1}. An alternative approach in understanding the dark energy arises from the holographic 
principle which states that the number of degrees of freedom related directly to entropy which scales with the enclosing area of the system. On 
the basis of the cosmological state of the holographic principle proposed by Fischler and Susskind \cite{Fischler1,Susskind1}, 
holographic model of dark energy has been proposed and studied widely in the literature \cite{Li1,Gong1,Wang1,Setare1,Gao1,Zhang1,Yang1}. 
The different dark energy models were mainly varied in the choice of the length scale $L$ appearing the dark energy density, 
$\rho_{de}=3 M_{pl} L^{-2}.$ One of the recent models of dark energy is holographic Ricci dark energy model, where the IR length scale 
$L$ is taken as the average radius of Ricci scalar curvature, $R^{-1/2}$ \cite{Li1,Gao1,Granda1}. This model has the following advantages.
First, the fine tuning problem can an be avoided in this model. Secondly, the presence of event horizon is not presumed in this model, so 
the causality problem be avoided. Moreover the coincidence problem can also be solved effectively in this model. Recently a form of modified 
holographic Ricci dark energy model (MHRDE) was proposed \cite{Chimento1} in connection with interaction between dark matter and  
dark energy. Detailed analysis of this model were found in the references \cite{Chimento2,Chatto1,tkm1}. Our study in this letter is with
this model of modified holographic Ricci dark energy interacting with the dark mater in the flat FRW universe. Interaction models where the dark energy 
weakly interacts with dark matter will throw light on how the energy flow between the dark sectors takes place as the universe evolves. It 
is often assumed that the decay rate from one sector to the other is proportional to the Hubble parameter for a good fit to the expansion 
history of the universe as determined by the supernova and CMB data \cite{Berger1}.

Motivated by the discovery of the black hole thermodynamics in the seventies, physicist have tried to analyses the thermodynamics of the 
cosmological models in an accelerated expanding universe \cite{Davies1,Huang1,Izquierdo1}. It was Bekenstein and Hawking who proved that 
the entropy of the black hole is proportional to the area of it's event horizon \cite{Bekenstein1,Hawking1}. By considering the entropy
decrease of a black hole during it's evaporation, the second law of thermodynamics was modified in such a way that the entropy of 
the black hole 
together with the entropy exterior to the blackhole is always increasing, and is known as the generalized second law (GSL). 
The thermodynamic properties of horizons were shown to exist in a more basic level by recasting the Einstein's field equation in the 
form of the fist law of thermodynamics \cite{Padma1,Padma2}. Bakenstein's idea was extended to cosmology by Gibbons and Hawking and have 
shown that cosmological event horizon also does possess entropy proportional to the horizon area \cite{Gibbons1}. They have shown it for 
de Sitter universe for which both apparent horizon and event horizon were coincide with each other. Since then many were studied about 
the validity of GSL in both apparent and event horizons \cite{Pollock1,Brustein1,Pavan1}. In the cosmological scenario the GSL means that 
the rate of change of entropy of the horizon together with that of the fluid within the horizon will always greater than equal to zero.
\begin{equation}
 {dS_{horizon} \over dt}+ {dS_{fluid} \over dt} \geq 0
\end{equation}
For interacting holographic dark energy model, Setare \cite{Setare2} have derived the constraint on the deceleration parameter,
for the fullfilment of the GSL in the non-flat universe enclosed by 
the event horizon. The first law of thermodynamics to the apparent horizon in FRW universe has been studied in \cite{Cai1,Bousso1}. In 
references \cite{Setare1,Izquierdo1} the validity of the GSL has been studied on the event horizon of a universe driven dark energy.
Using a
specific model of dark energy, Wang et. al \cite{Wang2} have proved that the GSL is obeyed at the apparent horizon of the universe but not on the 
event horizon. Validity of the GSL was studied for new holographic dark energy in reference \cite{Ujjal1} and have shown that, it is 
valid on the apparent horizon of the universe but partially valid on the event horizon of the universe. In reference \cite{Zhou1}, it has
been argued that, in contrast to the apparent horizon, the GSL of thermodynamics breakdown if one consider the universe is to be enveloped by 
the event horizon. These studies favors the apparent horizon as the physical boundary of the universe as far as the thermodynamics is 
concerned. The GSL of thermodynamics has also 
been studied in the framework of Braneworld \cite{Sheykhi1,Sheykhi2} and more general Levelock gravity \cite{Akbar1}.

At present there exist no unique theory which explaining the dark energy fully and since the modified holographic Ricci dark energy (MHRDE) is explaining the current acceleration of the universe without the 
assistance of fine tuning also explaining the coincidence problem, it is fair to analyze the thermodynamics of the universe with 
MHRDE. In this work we are examining the GSL of thermodynamics in a universe driven by MHRDE in interaction with dark matter and trying to 
answer the question whether Interacting MHRDE model favors apparent horizon or/and event horizon as the thermodynamic boundary. The 
paper is organized as follows. In section 2, we are presenting the relevant characteristics of the interacting MHRDE model. In section 3, we are 
presenting our analysis on the GSL under thermal equilibrium conditions both on the apparent horizon and event horizon of the universe. 
In section 4, we are briefly analyzing the conditions imposed by the validity of the GSL at the apparent horizon under non-equilibrium 
conditions. In section 5 we are presenting the conclusions.

   
\section{Interacting MHRDE model}
\label{sec2}

\begin{figure}[ht]
\centering
\includegraphics[scale=0.8]{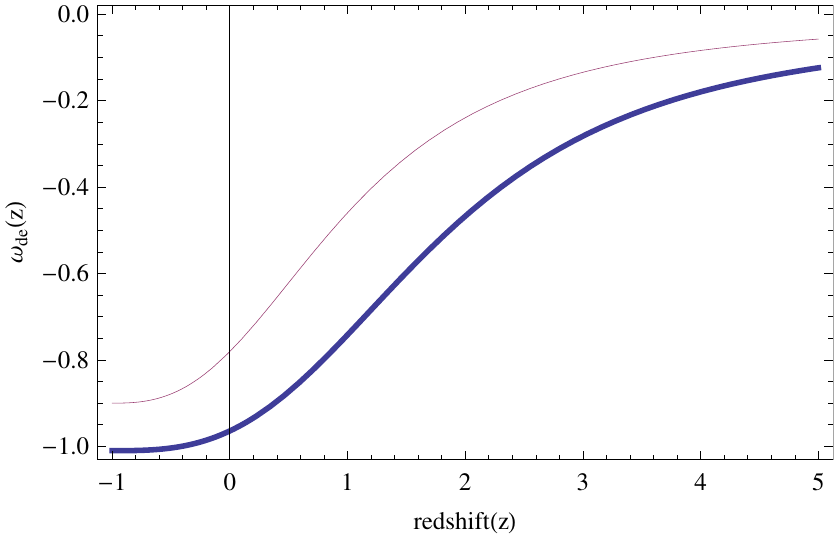}
\caption{Evolution of the equation of state of interacting MHRDE with redshift $z$  for interaction coupling term $b$=0.003 and parameter$(\alpha, \beta)$=(1.01,-0.01)-thick line,
 (1.2,0.1)-thin line.}
\label{fig:eosp}
\end{figure}

\begin{figure}[ht]
\centering
\includegraphics[scale=0.8]{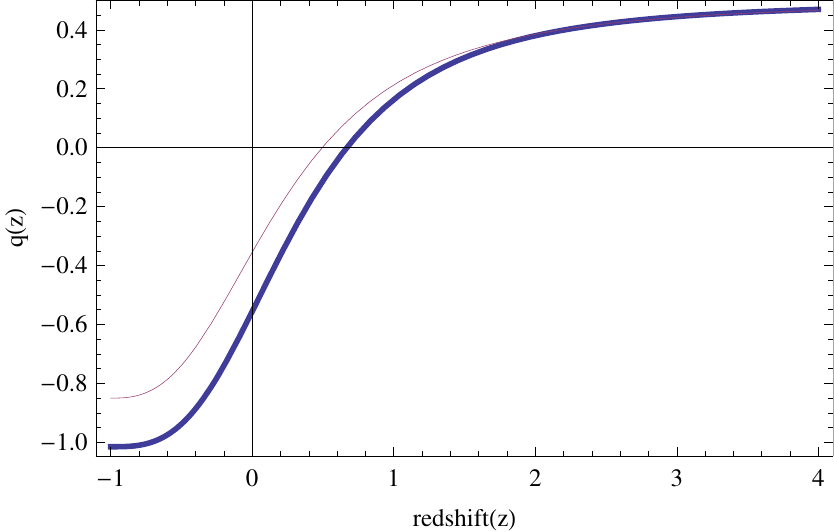}
\caption{Evolution of the $q$-parameter of interacting MHRDE with redshift. The plots for $(\alpha,\beta)$=(1.01,-0.01)-thick line,
 (1.2,0.1)-thin line with coupling constant $b=0.003.$}
\label{fig:q}
\end{figure}

We consider a flat FRW universe described by the equation,
\begin{equation}
 H^2 = \frac{1}{3}\left(\rho_{de} + \rho_m \right)
\end{equation}
where $H$ is the Hubble parameter, $\rho_{de}$ is the dark energy density and $\rho_m$ is dark matter density. The 
dark energy $\rho_{de}$ is taken as the modified holographic Ricci dark energy(where the Ricci scalar is taken as the IR cutt-off)
\cite{Chimento1,Chimento2},
\begin{equation}
 \rho_{de} = {2 \over (\alpha - \beta)} \left(\dot{H} + \frac{3\alpha}{2} H^2 \right)
\end{equation}
where $\dot{H}$ is the derivative of the $H$ with respect to cosmic time, $\alpha$ and $\beta$ are model parameters. The 
interaction between the dark energy and dark matter can be included through the continuity equations,
\begin{equation} \label{eqn:continuity1}
 \dot{\rho}_{de} + 3 H \left( \rho_{de} + p_{de} \right)= -Q
\end{equation}
\begin{equation} \label{eqn:continuity2}
 \dot{\rho}_m + 3 H \rho_m = Q
\end{equation}
Where $p_{de}$ is the pressure density of dark energy, dark matter is assumed to be pressure less, $Q$ is the interaction term and 
 over-dot representing derivative with respect to
time. Since there is no conclusive theory for the microscopic origin of the interaction between the dark sectors, 
one has to assume the form of $Q$ phenomenologically. The interaction term is chosen as, $Q=3bH\rho_m,$ where $b$ is the coupling constant 
and $b>0$ means energy is transferring from dark energy to cold dark matter. From Friedmann equation the second order differential 
equation for the weighted Hubble parameter $h=H/H_0$ (where $H_0$ is present value of the Hubble parameter) can be obtained,
\begin{equation}
 {d^2h^2 \over dx^2} + 3 \left(\beta - b +1 \right) {dh^2 \over dx} + 9 \beta \left(1-b\right) h^2 =0
\end{equation}
where $x=\ln a$ with $a$ as the scale factor of the universe.
The solution for this can be obtained as,
\begin{equation} \label{eqn:h}
 h^2 = k_1 e^{-3\beta x} + k_2 e^{-3(1-b)x}
\end{equation}
where the constants $k_1$ and $k_2$ are evaluated using the initial conditions,
\begin{equation}
 h^2|_{x=0} = 1 ,  \, \, \, \, {dh^2 \over dx}|_{x=0} = 3 \Omega_{de0} (\alpha - \beta) - 3 \alpha
\end{equation}
where $\Omega_{de0}$ is the value of present dark energy density and is related to present matter density $\Omega_{m0}$ as $\Omega_{de0}=1-\Omega_{m0}$ for the 
flat universe. The constants $k_1$ and $k_2$ are evaluated as
\begin{equation}
 k_1 = {\Omega_{de0} (\alpha - \beta) - \alpha - b +1 \over1- \beta - b  },  \, \, \, \, \, \, \, \, \, \, k_2 = 1-k_1
\end{equation}
 Again using Friedmann equation the dark energy density parameter can be obtained as,
\begin{equation} \label{eqn:DEP}
 \Omega_{de} = k_1 e^{-3\beta x} + k_2 e^{3(b-1)x} - \Omega_{m0} e^{-3x}
\end{equation}
From this the dark energy equation of state parameter can be calculated as,
\begin{equation} \label{eqn:eos}
 \omega_{de}= -1 + \left[ { k_1 \beta e^{-3\beta x} + k_2 (1-b) e^{-3(1-b)x} - \Omega_{m0} e^{-3x} \over 
k_1 e^{-3\beta x} + k_2 e^{-3(1-b)x} - \Omega_{m0} e^{-3x} } \right]
\end{equation}
In the non-interaction case, with $b=0$ and with $\Omega_{de0}=1,$ the coefficients, $k_1=1$ and $k_2=0$ and the equation of 
state reduces to the standard form, $\omega_{de}=-1+\beta,$ confirming the 
earlier observations \cite{tkm1} in this regard. The value of the interaction coupling constant, is chosen 
as $b=0.003$ in the present case, at which the model 
explaining the coincidence problem very well \cite{tkm2}. A sample of the evolution of the equation of state parameter 
with redshift $z$ ( note that $(1+z)=e^{-x}$ ) is shown figure in \ref{fig:eosp} for the 
model parameters \cite{Chimento1,Chimento2,tkm2}
$(\alpha,\beta)$=(1.01,-0.01),(1.2,0.1).
The figure shows that, for the parameters (1.01,-0.01) the equation of 
state parameter is varying from zero at very large values of redshift and evolves to $\omega_{de} \rightarrow -1$ as $z \rightarrow -1.$
 The present value of 
$\omega_{de}$ is found to around -0.97 from the figure, which is very close to the WMAP value -0.93 \cite{Komatsu1}.
as  $z\rightarrow -1,$ while for (1.2,0.1) the equation of state parameter $\omega_{de}>-1$ as $z\rightarrow -1.$ 
The deceleration parameter $q$ is found to be of the form
\begin{equation} \label{eqn:qp}
 q= -1 + \frac{3}{2} \left( { k_1 \beta e^{-3\beta x} - k_2 (b-1) e^{-3(1-b)x} \over k_1 e^{-3 \beta x} + k_2 e^{-3(1-b) x} } 
\right)
\end{equation}
The evolution of the $q$ parameter with redshift is shown in figure \ref{fig:q} for the same set of model parameters.
It is clear from the plot that the $q$ parameter starts with value around 0.5 at large redshift and evolves to negative values,
implying that the accelerated expansion commences in the recent past. For the parameters (1.01,-0.01) the transition to the accelerating phase at which $q$ parameter become 
negative, is occurred at redshift $z_T=0.68$, which very well in the observationally deducted range, $Z_T = 0.45 -0.70$ \cite{Alam2}.
The present value of the deceleration parameters from the plot is $q_0$=-0.57 is very much near to the observational 
prediction -0.60 \cite{Komatsu1}.

\section{GSL with thermal equilibrium conditions}

Here we will analyze the validity of the generalized second law of thermodynamics both at the apparent horizon and event horizon as the 
boundary of the universe. We consider a flat FRW universe, filled with dark energy 
and dark matter with mutual interaction. According to GSL the entropy of the fluid 
within the horizon and entropy of the horizon together do not decrease with time \cite{Karami1,tkm3}. In analyzing the 
validity of GSL we assume that
both the dark sectors are in equilibrium with the horizon, hence the temperature of both dark energy and dark matter are same as the 
temperature of the horizon. Such a thermal equilibrium is inevitable between the dark sectors because of the interaction between 
them. The dark sectors will soon come to equilibrium with the horizon. As a result the temperatures of all the components of the universe, 
will finally become equal to that of the the horizon temperature, $T_h = 1/2\pi R_h$ \cite{Gibbons1,Cai2}, where $R_h$ is radius of the 
horizon. Further we will also assume that the first law of thermodynamics given by
\begin{equation} \label{eqn:FL}
 T dS = dE + P dV 
\end{equation}
is satisfied at the horizon (both apparent and event horizons) where
$T, S, E, P , V$ are the temperature, entropy, internal energy, pressure and volume of the system.

\begin{figure}[ht]
\centering
\includegraphics[scale=0.75]{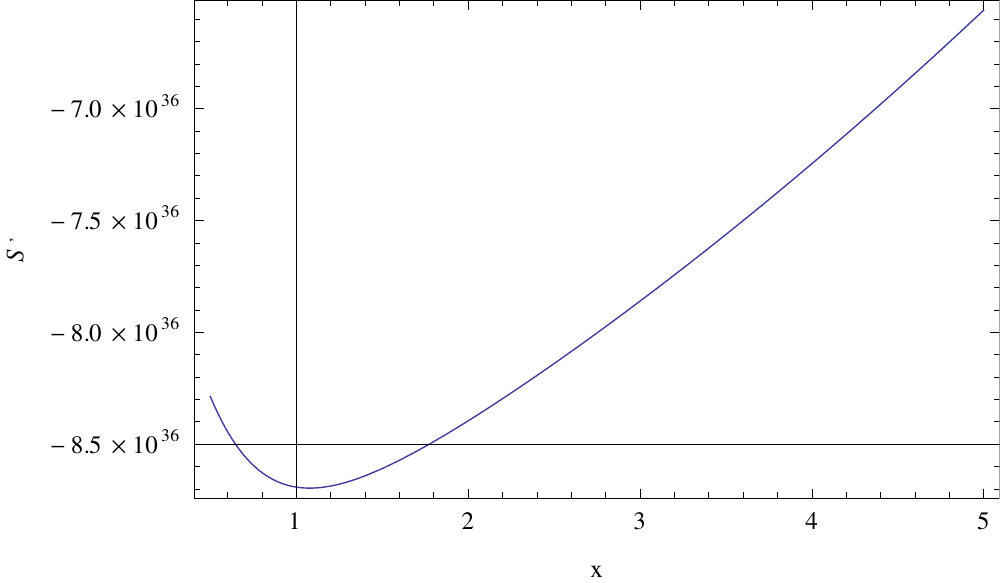}
\caption{Plot of the rate of change of the total entropy $S^{'}$ with $x$ for the model parameter $(\alpha,\beta)=(1.01,-0.01)$ with 
interaction coupling constant $b=0.003.$}
\label{fig:gsl1}
\end{figure}

We will analyze first the validity of the GSL at the apparent horizon, with horizon radius $R_A = 1/H$ which is also same as the Hubble horizon for 
the flat universe. From the above equation the rate of change of entropy of the dark energy and dark matter with 
respect to $x=\ln a$ are
\begin{equation} \label{eqn:ER}
 S^{'}_{de} = {E^{'}_{de} + P_{de} V^{'} \over T},  \, \, \, \, \, S^{'}_m = {E^{'}_m \over T}
\end{equation}
here the pressure due to dark matter is taken as zero and $T=T_h$ under equilibrium conditions. The volume and temperature are given as 
\begin{equation}\label{eqn:VT}
 V = {4\pi^3 \over 3H^3}, \, \, \, \, \, T_h={H \over 2\pi}
\end{equation}
The internal energies of the dark energy and dark matter are 
\begin{equation} \label{eqn:energy}
 E_{de} = {4\pi \rho_{de} \over 3 H^3}, \, \, \, \, \, E_m = {4\pi \rho_m \over 3 H^3}
\end{equation}
The entropy of the apparent horizon is $S_A = \pi/H^2,$ so the derivative with respect $x$ is 
\begin{equation} \label{eqn:EH}
 S^{'}_A = {-2\pi H^{'} \over H^3}
\end{equation}
From equations (\ref{eqn:ER}-\ref{eqn:EH}), the rate of change in total entropy with respect to $x$ is 
\begin{equation}
 S^{'}=S^{'}_{de}+S^{'}_m+S^{'}_A = {2\pi \over H^2}+{2\pi \over H^2} \left(1+12\pi (1+\omega_{de} \Omega_{de}) \right) (0.5 + 1.5 \omega_{de}
\Omega_{de})
\end{equation}
where we used the relation for the $q-$parameter for the flat universe,
\begin{equation}
 q=-1-{\dot{H} \over H^2} = \frac{1}{2}\left(1+3\omega_{de} \Omega_{de} \right)
\end{equation}
where the dot represents the derivative with respect to time. In figure \ref{fig:gsl1}, we have plotted the evolution of the rate 
\begin{figure}[ht]
\centering
\includegraphics[scale=0.75]{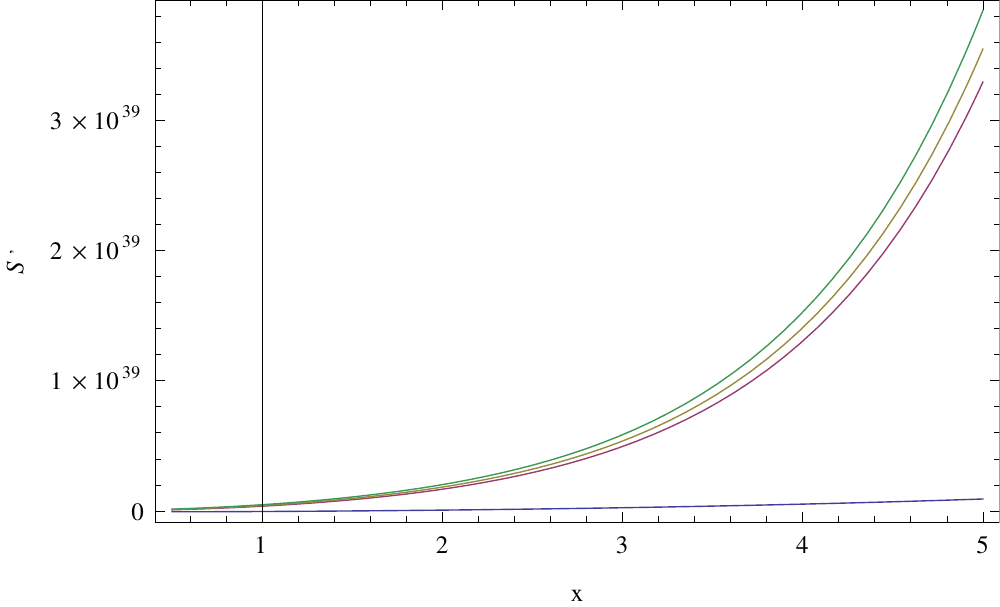}
\caption{Plot of the rate of change of the total entropy $S^{'}$ with $x$ for model parameter $(\alpha,\beta)$=(1.2,0.1)-first 
line from the bottom, (1.2,0.3)-second line from the bottom, (1.3,0.3)-third line from the bottom, (1.4,0.3)-topmost line with 
$b$=0.003.}
\label{fig:gsl2}
\end{figure}
of the total entropy, $S^{'}$ with $x$ using equations (\ref{eqn:h}),(\ref{eqn:DEP}) and (\ref{eqn:eos}) for the model parameters 
$(\alpha,\beta)=(1.01,-0.01).$ This figure shows that the variation of 
the total entropy is always negative which implies that the GSL of thermodynamics
is not satisfied at the apparent horizon for the model parameters (1.01,-0.01). Next we will check the status of GSL for the 
model parameters $(1.2,0.1),(1.2,0.3),(1.3,0.3),(1.4,0.3)$, and the respective rate of the total 
entropy with $x$ is given in figure \ref{fig:gsl2}. The figure shows that the rate of total entropy is positive throughout the evolution 
of the universe, hence the GSL is satisfied always at 
apparent horizon for these parameters with MHRDE interacting with matter within the apparent horizon. This shows that the apparent 
horizon can be a good boundary for studying the cosmology and thermodynamics of the universe with interacting MHRDE. For parameter 
(1.01,-0.01), the equation of state $\omega_{de} \rightarrow -1$, while for other parameters the $\omega_{de} > -1$ through out the expansion 
of the universe. This indicate that apparent horizon of the universe with interacting MHRDE satisfies the GSL of thermodynamics as 
far as $\omega_{de} > -1.$ This limit on the equation of state parameter otherwise indicating that the universe is not entering the phantom 
phase of expansion at any stage of it's evolution as far as $\omega_{de}> -1.$. So it can be concluded that the universe consisting of 
interacting MHRDE satisfy the 
GSL at the apparent horizon for model parameters for which the universe doesn't  go in to the phantom phase.

\begin{figure}[ht]
\centering
\includegraphics[scale=0.75]{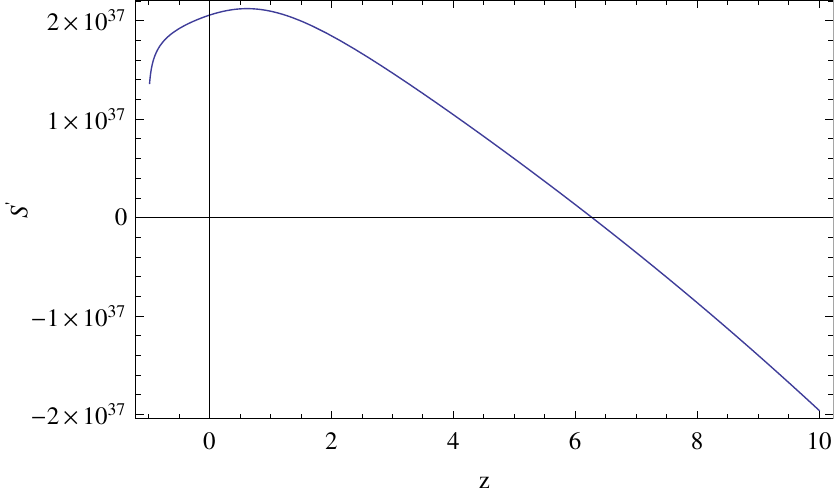}
 \caption{The plot of $S^{'}$ versus x for the model parameters $(\alpha,\beta)$=(1.01,-0.01) at the event horizon of the universe
with interaction coupling term $b$=0.003.}
\label{fig:gsl3}
\end{figure}

Next we consider the validity of GSL at the event horizon of the universe with interacting MHRDE. The rate of change of the entropy can 
be calculated using the relations (\ref{eqn:ER}). For which the volume and temperature are,
\begin{equation}
 V = {4\pi \over 3}R_E^3,  \, \, \, \, \, \, \, \, T_E={1 \over 2\pi R_E}.
\end{equation}
where $R_E$ is the event horizon radius of the universe.
The event horizon distance of a universe 
can be obtained using the relation \cite{Ujjal1}
\begin{equation}
R_E = a(t) \int_t^{\infty} {dt^{'} \over  a(t^{'})} = -{1 \over (1+z) H_0} \int_z^{-1} {dz^{'} \over h}
\end{equation}
where $z$ is the redshift. With the equation (\ref{eqn:h}), the event horizon radius is become
\begin{center}
\begin{eqnarray} 
 R_E = ({\gamma \over H_0}) {1 \over \sqrt{k_2 (1+z)^{3(1-b)} }} \times \, \, \, \, \, \, \, \, \, \, \, \, \, \, \, \, \, \, \, \, \, 
\, \, \, \, \, \, \, \, \, \, \, \, \, \, \, \, \, \, \, \, \, \, \, \, \, \, \, \, \, \, \, \, \, \, \, \, \, \, \, \, 
\nonumber \\  {_2}F_1[0.5, -{0.5 \over 3\beta-3(1-b)}, 1-{0.5 \over 3\beta-3(1-b)}, -{k_1
\over k_2} (1+z)^{3\beta-3(1-b)}]
\end{eqnarray}
\end{center}
where $\gamma = 2.0182,$ a constant and ${_2}F_1$ is the 
hypergeometric function. Using equation (\ref{eqn:ER}), the sum of the entropies of dark energy and dark matter can be obtained as,
\begin{equation}
 S_{de}^{'} + S_m^{'} = -24\pi^2 \left(1+\omega_{de} \Omega_{de} \right)HR_E^3
\end{equation}
Adding the entropy of the event horizon and using the standard relation $\dot{H}=HR_E -1$ \cite{Karami2}, the rate of the total entropy can
be obtained as,
\begin{equation}
 S^{'}= S^{'}_{de}+S^{'}_m+S^{'}_E = -24\pi^2\left(1+\omega_{de} \Omega_{de} \right)HR_E^3+2\pi R_E^2-2\pi H^{-1}R_E
\end{equation}
which shows that the GSL is satisfied, i.e. $S^{'} \geq 0$ if
\begin{equation}
 \omega_{de} \Omega_{de} \leq -1+\frac{1}{12\pi H R_E} {d\log R_E \over dx} 
\end{equation}
This shows that as far as $R_E^{'} \geq 0,$ the equation of state $\omega_{de}$ is always less than -1 in a dark energy dominated 
universe, but on the other hand if $R_E^{'} \leq 0$ $\omega_{de} < -1$ always. In figure \ref{fig:gsl3} we plot $S^{'}$ for the parameter set $(\alpha,
\beta)$=(1.01,-0.01), and it is showing that the GSL is satisfied at the event horizon only partially. The GSL is satisfied only in the 
remote past and far future in the evolution.
\begin{figure}[ht]
 \centering
\includegraphics[scale=0.75]{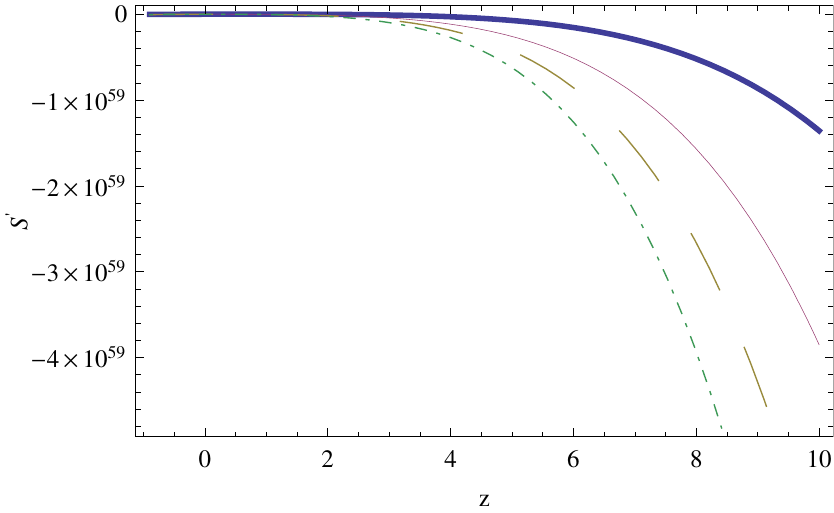}
\caption{Plot of $S^{'}$ with $z$ for model parameters $(\alpha,\beta)$=(1.2,0.1)-thick continuous line, (1.2,0.3)-thin continuous line,
(1.3,0.3)-dashed line,(1.4,0.3)-dot-dashed line, with interaction coupling term $b$=0.003.}
\label{fig:gsl4}
\end{figure}
For other model parameters parameters it is clear form figure \ref{fig:gsl4} that the GSL is not satisfied
at the event horizon at large redshifts. But when the redshift is less than 2, the GSL is found satisfy with limiting 
condition that $S{'}=0.$ So it can be concluded that the GSL is not satisfied through out the evolution of the universe. 
These results shows that the event horizon of the universe with interacting MHRDE cannot be taken as physical boundary for the thermodynamics 
of the universe.

\section{GSL with thermal non-equilibrium conditions}

\begin{figure}[ht]
 \centering
\includegraphics[scale=0.75]{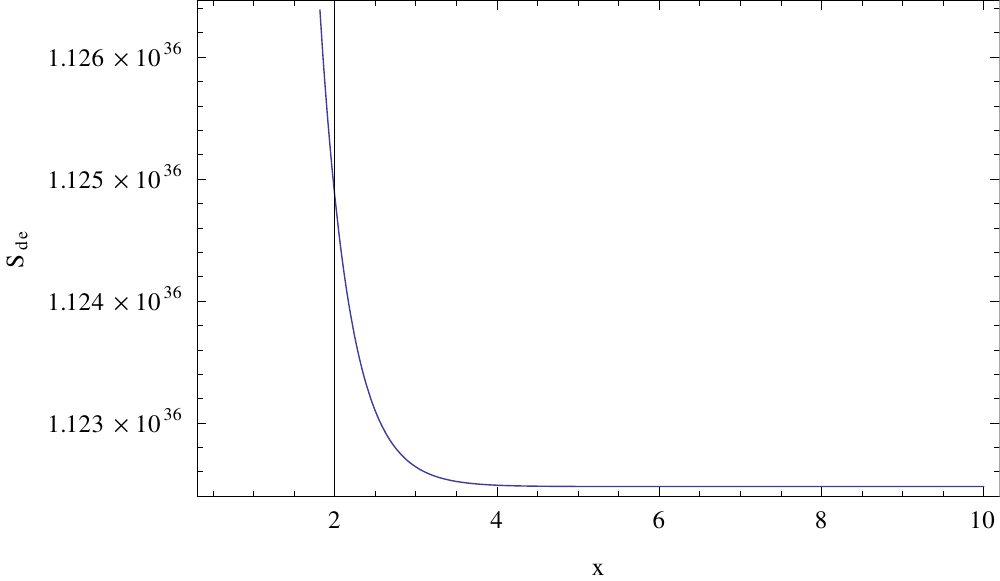}
\caption{Evolution of the dark energy entropy with $x$ for the model parameters (1.2,0.1) }
\label{fig:darkentro}
\end{figure}

\begin{figure}[ht]
\centering
\includegraphics[scale=0.75]{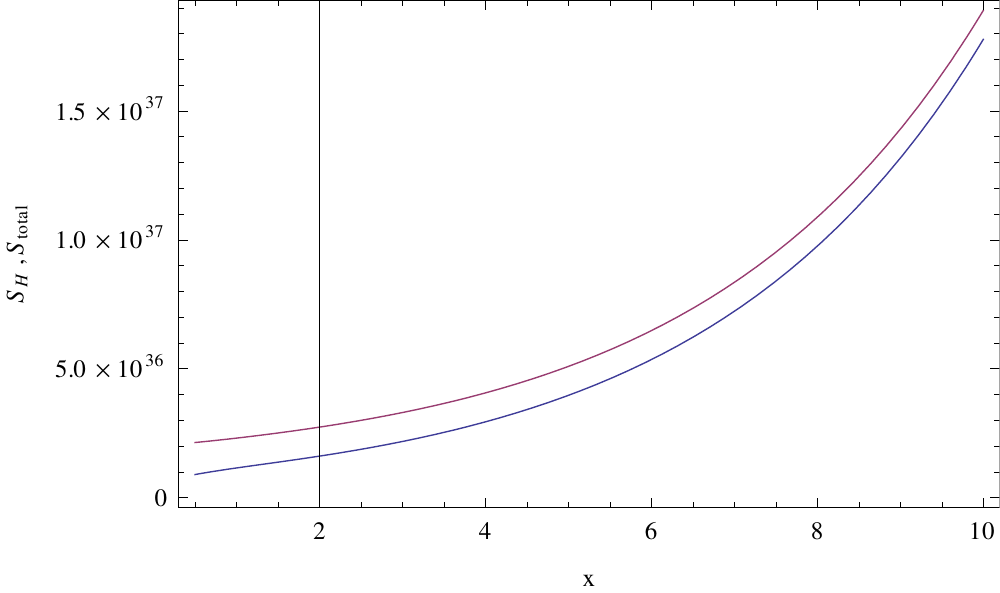}
\caption{Evolution of the entropy of the apparent horizon with $x-$(bottom line) and total entropy with $x-$(top line) for the model 
parameters (1.2,0.1).}
\label{fig:totalentro}
\end{figure}

Here will analyze the validity of the GSL by assuming that, the different components are not in thermal equilibrium. So the dark 
energy and dark matter were evolved with different temperature and the horizon have a still different temperature. This is strictly 
possible when there is no interaction between the components of the universe. So we will take $Q=0$ in the conservation equations 
\ref{eqn:continuity1} and \ref{eqn:continuity2}. The rate of change of the total entropy of the universe enclosed by the 
apparent horizon can be then obtained as
\begin{equation}
 S^{'} = {12\pi \over H} {\Omega_{de} \over T_{de}} (1+\omega_{de}) q + {12\pi \over H} {\Omega_m \over T_m} q - {2\pi H^{'} \over H^3}
\end{equation}
where $T_{de}$ and $T_m$ are the temperatures of the dark energy and dark matter respectively. The validity of the GSL requires that 
$S^{'} > 0,$ which gives a constraint on the temperature of the dark energy in a universe dominated with dark energy as
\begin{equation}
 T_{de} \geq -\left({6 q \over 1+q} \right)(1+ \omega_{de}) H
\end{equation}
By taking the observational value of $q_0$ and $\omega_{de0}$ from WMAP results, -0.60 and -0.93 respectively, then it can be seen that
$T_{de0}>H_0.$ This inevitably shows that the temperature of the dark energy is greater than that of the event horizon (taking event 
horizon temperature as $T_h \sim H $). As a result of this, there occurs a flow of dark energy 
towards the apparent horizon. In literature there are examples for the anticipation of the flow of the fluid from within the horizon towards 
the horizon \cite{Davies2,tkm3}.
This flow towards the apparent horizon would possibly lead to the crossing of the dark energy across the horizon.
Hence the entropy of the dark energy will decrease as the universe evolves. But the validity of the GSL in turn requires that, the 
decrease in the dark energy entropy is balanced by the increase in the horizon entropy such that the total entropy will never decrease. 
Consider a universe dominated by dark energy. The 
entropy of the dark energy can be obtained from simple considerations as \cite{Kolb1},
\begin{equation}
 S_{de}= \left({\rho_{de} + p_{de} \over T_{de}} \right) V = {8\pi^2 (1+\omega_{de}) \Omega_{de} \over c H^2}
\end{equation}
where we have taken the temperature of the dark energy is proportional to the horizon temperature as, $T_{de} = c \frac{H}{2\pi}$ 
\cite{Izquierdo1}, $c$ is a numerical constant greater than one to 
ensure that the dark energy temperature is greater than that of the horizon. The entropy of the apparent horizon is,
\begin{equation}
 S_H = {\pi \over H^2}
\end{equation}
Using the expressions (\ref{eqn:h}), (\ref{eqn:DEP}) and (\ref{eqn:eos}), we have plotted $S_{de}$ and $S_H \& S_{total}$ for $c=1.25$ 
and are shown in figures
\ref{fig:darkentro} and \ref{fig:totalentro}, where $S_{total}=S_H+S_{de}.$
It is evident the figures that the entropy of the dark energy is decreasing, the entropy of the horizon is increasing and the total 
entropy comprising the entropies of the dark energy and horizon is also increasing as result the GSL is satisfied.

\section{Conclusions}

In this work we have considered the flat FRW universe filled with holographic Ricci dark energy and dark matter and 
analyzed the validity of the GSL of thermodynamics at the apparent and event horizons of the universe. We have done the 
analysis for both thermal equilibrium and non-equilibrium conditions.

 We have first analyzed the validity of the GSL
under thermal equilibrium conditions at which the temperature of the dark energy and dark matter were equal to the temperature 
of the horizon. 
The thermal equilibrium is achieved mainly through the interaction between the dark energy and dark matter.
 We have found that, with modified holographic Ricci dark energy, the GSL is 
satisfied with apparent horizon as the boundary of the universe for all model parameters except 
(1.01,-0.01). For the parameters (1.01,-0.01), 
there exist a possibility for the universe to
go into the phantom phase of expansion which seems to causes the partial violation of the GSL.  
Hence the apparent horizon can be taken as the physical boundary for thermodynamic considerations as far as the 
universe doesnot entering the phantom phase of expansion or equivalently $\omega_{de} < -1.$ This conclusion is in agreement with 
many other works \cite{Pollock1,Brustein1,Pavan1,Setare1,Wang1,Ujjal1,Zhou1} with different forms of dark energies.
In the case of the universe with event horizon as the boundary, we have found that the GSL is partially satisfied for all the parameters 
including (1.01,-0.01). For the parameters (1.01,-0.01) the GSL is satisfied at the recent past and recent future and violated in other 
regions. But for other parameters, it is clear form figure \ref{fig:gsl3} that, GSL is not satisfied in the past 
corresponding to redshift $z>2.$ For $z<2$, GSL is satisfied for all 
the model parameters with the condition $\displaystyle {dS/dt} = 0.$ Because of this partial fulfillment of the GSL at the 
event horizon, it cannot be taken 
as a physical boundary of the universe. This is also in tune with conclusion with other models of dark energy as noted in the reference
cited above.

Under non-equilibrium condition we have obtained the constraints imposed by the validity of the GSL at the apparent horizon. 
In this case the dark energy and dark matter are 
non-interacting, hence their temperatures are not equal and are different from the horizon temperature. We have found 
that for the GSL is to be satisfied the temperature of the 
dark energy must be greater than that of the apparent horizon. This implies that the dark energy can flow towards the horizon and even 
cross the horizon. As a result the entropy of the dark energy must decrease. On the other hand the entropy of the horizon and also 
the total entropy must increase which ensures the validity of the GSL. Assuming the temperature of dark energy, $T_{de}= c (H/2\pi)$, 
with $c>1,$ we have calculated the entropies of dark energy and horizon. The evolution of these entropies, figures \ref{fig:darkentro} and
\ref{fig:totalentro}, shows that the entropy of the dark is deceasing but the entropy of the horizon is increasing. The increase in the 
horizon entropy is compensating the decrease in the entropy of the dark energy, so that the total entropy is always increasing.


\begin{thebibliography}{00}

\bibitem{PM1} S. Perlmutter et. al., Astrophys. J.517 (1999) 565.
\bibitem{AG1} A. G. Riess et. al. Astron. J. 116 (1998) 1009.
\bibitem{Garno1} P. M. Garnovich et. al. Astrophys. J. 493 (1998) L53.
\bibitem{Spergel1} Spergel D. N. et. al. Astrophys. j. Suppl. Ser. 148 (20030 175.
\bibitem{Ratra1} B. Ratra and P.J.E. Peebles, Phys. Rev.D 37 (1988) 3406.
\bibitem{Wetterich1} C. Wetterich, Nucl. Phys. B 302 (1988) 302.
\bibitem{Zlatev1} I. Zlatev, L. Wang and P.J. Steinhardt, Phys. Rev. Lett. 82 (1999) 896.
\bibitem{Chiba1} T. Chiba, T. Okabe and M. Yamaguchi, Phys. Rev. D 62 (2000) 023511.
\bibitem{Kamenshchik1} A. Kamenshchik, U. Moschella and V. Pasquier, Phys. Lett. B 511 (2001) 265.
\bibitem{Alam1} U. Alam, V. Sahni, T. D. Saini and A. A. Starobinsky, Mon. Not. Roy. Astron. Soc. 344 (2003) 1057.
\bibitem{Caldwel1} R. R. Caldwell, Phys. Lett. B 545 (2002) 23.
\bibitem{Feng1} B. Feng, X. L. wang and X. M. Zhang, Phys. Lett. B 607 (2005) 35.
\bibitem{Fischler1} W. Fischler and L. Susskind, hep-tp/9806039.
\bibitem{Susskind1} L. Susskind, J. Math. Phys. 36 (1995) 6377.
\bibitem{Li1} M. Li, Phys. Lett. B 603 (2004) 1
\bibitem{Gong1} Y. Gong, Phys. Rev. D 70 (2004) 064029.
\bibitem{Wang1} B. Wang, E. Abdalla and R. K. Su, Phys. Lett. B 611 (2005) 
\bibitem{Setare1} M R Setare, Phys. Lett. B 641 (20060 130.
\bibitem{Gao1} C. Gao, F. Q. Wu, X. Cheng and Y. -G. Shen, Phys. Rev. D 79 (2009) 043511.
\bibitem{Zhang1} L. Zhang, P. Wu and H. Yu, Eur. Phys. J. C 71 (2011) 1588.
\bibitem{Yang1} R. -J. Yang, Z. -H. Zhu and F. Wu, Int. J. Mod. Phys. A 26 (2011) 317.
\bibitem{Granda1} L. N. Granda and A. Oilveros, Phys. Lett. B 671 (2009) 199.
\bibitem{Chimento1} L. P. Chimento and M. G. Richarte, Phys. Rev. D 81 (2010) 043525.
\bibitem{Chimento2} L. P. Chimento and M. G. Richarte, Phys. Rev. D 84 (2011) 123507.
\bibitem{Chatto1} S. Chattopadhayay, Eur. Phys. J. Plus 126 (2011) 130.
\bibitem{tkm1} Titus K. Mathew, Jishnu Suresh and Divya D. Nair, Int. J. Mod. Phys. D 22 (2013) 1350056.
\bibitem{Berger1} M. S. Berger and H. Shojaei, Phys. Rev. D 74 (2006) 043530.
\bibitem{Davies1} P. C. W. Davies Class. Quan. Grav. 5 (1988) 1349.
\bibitem{Huang1} Q. Huang and M. Li, JCAP 0408 (2004) 013.
\bibitem{Izquierdo1} G. Isquierdo and D. Pavon, Phys. Lett. B 633 (2006) 420.
\bibitem{Bekenstein1} J. D. Bekenstein, Phys. Rev. D 7 (1973) 2333.
\bibitem{Hawking1} S. W. Hawking, Commun. Math. Phys. 43 91975) 199.
\bibitem{Padma1} T. Padmanabhan, Phys. Rep. 49 (2005) 406.
\bibitem{Padma2} T. Padmanabhan, Phys. Rep. 73 (2010) 046901.
\bibitem{Gibbons1} G. W. Gibbons and S. W. Hawking, Phys. Rev. D 15 (1977) 2738.
\bibitem{Pollock1} M. D. Pollock and T. P. Singh, Class. Quantum Grav. 6 (1989) 901.
\bibitem{Brustein1} R. Brustein, Phys. Rev. Lett. 84 (2000) 2072.
\bibitem{Pavan1} D. Pavan, Class. Quantum Grav. 7 (1990) 487.
\bibitem{Setare2} M. R. Setare, JCAP 023 (2007) 0701.
\bibitem{Cai1} R. -G. Cai and P. K. Song, JHEP 02 (2005) 50
\bibitem{Bousso1} R. Bousso, Phys. Rev. D 71 (2005) 064024.
\bibitem{Wang2} B. Wang, Y. Gong and E. Abdalla, Phys. Rev. D 81 (2010) 023007.
\bibitem{Ujjal1} Ujjal Debnath and Surajit Chattopadhyay, Int. J. Theor. Phys. 52 (2013) 1250.
\bibitem{Zhou1} J. Zhou, B. Wang, Y. Gong and E. Abdalla, Phys. Lett. B 652 (2007) 86.
\bibitem{Sheykhi1} A. Sheykhi and B. Wang, Phys. Lett. B 678 (2009) 434.
\bibitem{Sheykhi2} A. Sheykhi and B. Wang, Mod. Phys. Lett. A 25 (2010) 1199.
\bibitem{Akbar1} M. Akbar, Int. J. Theor. Phys. 48 (2009) 2665.
\bibitem{tkm2} Praseetha P. and Titus K. Mathew, arXiv:1309.3136v1[astro-ph.CO]
\bibitem{Karami1} K. Karami and S. Ghaffari, Phys. Lett. B 688 (2010) 125.
\bibitem{tkm3} Titus K Mathew, Aiswarya R and Vidya K Soman, Eur. Phys. J C 73 (2013) 2619.
\bibitem{Cai2} R. -G. Cai, Cao and Y. P. Hu, Class. Quantum Grav. (2009) 26 155018.
\bibitem{Komatsu1} Komatsu E et. al., [WMAP Collaboration], Astrophys. J. Suppl. 192 (2011) 18.
\bibitem{Alam2} Alam U, Sahni V and A A Starobinsky, JCAP 406 (2004) 008.
\bibitem{Karami2} K. Karami, S. Ghaffari and M.M. Soltanzadeh, Class.Quant.Grav. 27 (2010) 205021. 
\bibitem{Davies2} T. M. Davies, P. C. W. Davies and Lineweaver, Class. Quantum Grav. 20 (2003) 2753.
\bibitem{Kolb1} Edward W. Kolb and Michael S. Turner, The Early Universe, Addison-Wesley, California, 1990.


\end{thebibliography}
\end{document}